\documentclass[12pt,preprint]{aastex}

\shorttitle{Asteroseismology of solar-type stars with
  K2}\shortauthors{Chaplin et al.}

\begin{document}

\title{Asteroseismology of solar-type stars with K2: detection of
  oscillations in C1 data}

\author{
W. J. Chaplin\altaffilmark{1,2},
M. N. Lund\altaffilmark{2,1},
R. Handberg\altaffilmark{2,1},
S. Basu\altaffilmark{3},
L. A. Buchhave\altaffilmark{4,5},
T. L. Campante\altaffilmark{1,2},
G. R. Davies\altaffilmark{1,2},
D. Huber\altaffilmark{6,7,2},
D. W. Latham\altaffilmark{4},
C. A. Latham\altaffilmark{4},
A. Serenelli\altaffilmark{8},
H. M. Antia\altaffilmark{9},
T. Appourchaux\altaffilmark{10},
W. H. Ball\altaffilmark{11},
O. Benomar\altaffilmark{12},
L. Casagrande\altaffilmark{13},
J. Christensen-Dalsgaard\altaffilmark{2},
H. R. Coelho\altaffilmark{1,2},
O. L. Creevey\altaffilmark{14},
Y. Elsworth\altaffilmark{1,2},
R. A. Garc\'ia\altaffilmark{15},
P. Gaulme\altaffilmark{16},
S. Hekker\altaffilmark{17,2},
T. Kallinger\altaffilmark{18},
C. Karoff\altaffilmark{2},
S. D. Kawaler\altaffilmark{19},
H. Kjeldsen\altaffilmark{2},
M. S. Lundkvist\altaffilmark{2},
F. Marcadon\altaffilmark{10},
S. Mathur\altaffilmark{20},
A. Miglio\altaffilmark{1,2},
B. Mosser\altaffilmark{21},
C. R\'egulo\altaffilmark{22},
I. W. Roxburgh\altaffilmark{23},
V. Silva Aguirre\altaffilmark{2},
D. Stello\altaffilmark{6,2},
K. Verma\altaffilmark{9},
T. R. White\altaffilmark{11},
T. R. Bedding\altaffilmark{6,2},
T. Barclay\altaffilmark{24,25},
D. L. Buzasi\altaffilmark{26},
S. Deheuvels\altaffilmark{27},
L. Gizon\altaffilmark{17,11,28},
G. Houdek\altaffilmark{2},
S. B. Howell\altaffilmark{24},
D. Salabert\altaffilmark{15},
D. R. Soderblom\altaffilmark{29}
}

\altaffiltext{1}{School of Physics and Astronomy, University of
  Birmingham, Edgbaston, Birmingham, B15 2TT, UK}

\altaffiltext{2}{Stellar Astrophysics Centre (SAC), Department of
  Physics and Astronomy, Aarhus University, Ny Munkegade 120, DK-8000
  Aarhus C, Denmark}

\altaffiltext{3}{Department and Astronomy, Yale University, New Haven,
  CT, 06520, USA}

\altaffiltext{4}{Harvard-Smithsonian Center for Astrophysics, 60
Garden Street Cambridge, MA 02138 USA}

\altaffiltext{5}{Centre for Star and Planet Formation, Natural History
Museum of Denmark, University of Copenhagen, DK-1350 Copenhagen,
Denmark}

\altaffiltext{6}{Sydney Institute for Astronomy, School of Physics,
  University of Sydney, Sydney, Australia}

\altaffiltext{7}{SETI Institute, 189 Bernardo Avenue, Mountain View,
  CA 94043, USA}

\altaffiltext{8}{Instituto de Ciencias del Espacio (ICE-CSIC/IEEC)
  Campus UAB, Carrer de Can Magrans, s/n 08193 Cerdanyola del
  Vall\'es, Spain}

\altaffiltext{9}{Tata Institute of Fundamental Research, Homi Bhabha
Road, Mumbai 400005, India}

\altaffiltext{10}{Institut d'Astrophysique Spatiale, Universit\'e Paris
11, CNRS (UMR8617), Batiment 121, F-91405 Orsay Cedex, France}

\altaffiltext{11}{Institut f\"ur Astrophysik, Georg-August-Universit\"at
G\"ottingen, Friedrich-Hund-Platz 1, 37077, G\"ottingen, Germany}

\altaffiltext{12}{The University of Tokyo, Tokyo 113-0033, Japan}

\altaffiltext{13}{Research School of Astronomy \& Astrophysics,
Australian National University, Mt Stromlo Observatory, via Cotter Rd,
Weston, ACT 2611, Australia}

\altaffiltext{14}{Laboratoire Lagrange, Universit\'e de Nice
Sophia-Antipolis, UMR 7293, CNRS, Observatoire de la C\^ote d'Azur,
Nice, France}

\altaffiltext{15}{Laboratoire AIM Paris-Saclay, CEA/DSM - CNRS -
Univ. Paris Diderot - IRFU/SAp, Centre de Saclay, F-91191
Gif-sur-Yvette Cedex, France}

\altaffiltext{16}{Department of Astronomy, New Mexico State University,
P.O. Box 30001, MSC 4500, Las Cruces, NM 88003-8001, USA; Apache Point
Observatory, 2001 Apache Point Road, P.O. Box 59, Sunspot, NM 88349,
USA}

\altaffiltext{17}{Max-Planck-Institut f\"ur Sonnensystemforschung,
Justus-von-Liebig-Weg 3, 37077, G\"ottingen, Germany}

\altaffiltext{18}{Institut f\"ur Astronomie, Universit\"at Wien,
T\"urkenschanzstr. 17, 1180 Wien, Austria}

\altaffiltext{19}{Department of Physics and Astronomy, Iowa State
  University, Ames, IA 50011, USA}

\altaffiltext{20}{Space Science Institute, 4750 Walnut Street Suite 205,
Boulder CO 80301, USA}

\altaffiltext{21}{LESIA, Observatoire de Paris, PSL Research
  University, CNRS, Universit\'e Pierre et Marie Curie, Universit\'e
  Denis Diderot, 92195 Meudon, France}

\altaffiltext{22}{Instituto de Astrof\'isica de Canarias, 38205 La
Laguna, Tenerife, Spain; Universidad de La Laguna, Dpto. de
Astrof\'isica, 38206 La Laguna, Tenerife, Spain}

\altaffiltext{23}{Astronomy Unit, Queen Mary University of London, Mile
End Road, E1 4NS, London, UK}

\altaffiltext{24}{NASA Ames Research Center, Moffett Field, CA 94035,
  USA}

\altaffiltext{25}{Bay Area Environmental Research Inst., 560 Third St.,
  West Sonoma, CA 95476, USA}

\altaffiltext{26}{Department of Chemistry and Physics Florida Gulf Coast
  University 10501 FGCU Boulevard South Fort Myers, FL 33965-6501,
  USA}

\altaffiltext{27}{Universit\'e de Toulouse, UPS-OMP, IRAP, 31028,
  Toulouse, France}

\altaffiltext{28}{Center for Space Science, New York University Abu
  Dhabi, P.O. Box 129188, Abu Dhabi, UAE}

\altaffiltext{29}{Space Telescope Science Institute, Baltimore, MD
  21218 and Center for Astrophysical Sciences, Department of Physics
  and Astronomy, Johns Hopkins University, Baltimore, MD 21218, USA}

\begin{abstract}

We present the first detections by the NASA K2 Mission of oscillations
in solar-type stars, using short-cadence data collected during K2
Campaign\,1 (C1). We understand the asteroseismic detection thresholds
for C1-like levels of photometric performance, and we can detect
oscillations in subgiants having dominant oscillation frequencies
around $1000\,\rm \mu Hz$. Changes to the operation of the
fine-guidance sensors are expected to give significant improvements in
the high-frequency performance from C3 onwards. A reduction in the
excess high-frequency noise by a factor of two-and-a-half in amplitude
would bring main-sequence stars with dominant oscillation frequencies
as high as ${\simeq 2500}\,\rm \mu Hz$ into play as potential
asteroseismic targets for K2.

\end{abstract}


 \section{Introduction}
 \label{sec:intro}

Asteroseismology of solar-type stars has been one of the major
successes of the NASA \emph{Kepler} mission (Gilliland et
al. 2010a). The nominal mission provided data of exquisite quality for
unprecedented numbers of low-mass main-sequence stars and cool
subgiants. Asteroseismic detections were made in more than 600 field
stars (Chaplin et al. 2011a; 2014), including a sample of
\emph{Kepler} planet hosts (Huber et al. 2013).  These data have
enabled a range of detailed asteroseismic studies (see Chaplin \&
Miglio 2013 and references therein), many of which are ongoing.

The nominal mission ended in 2013 May with the loss of a second of the
spacecraft's four onboard reaction wheels. This meant the spacecraft
could no longer maintain three-axis stabilized pointing. However,
thanks to the ingenuity of the mission teams, \emph{Kepler} data
collection has continued as a new ecliptic-plane mission, K2 (Howell
et al. 2014). Targeting stars in the ecliptic minimizes the now
unconstrained roll about the spacecraft boresight, thereby helping to
compensate for the loss of full three-axis stability.  The degraded
photometric performance presents particular challenges for the
detection of oscillations in solar-type stars. The oscillations are
stochastically excited and intrinsically damped by near-surface
convection. While this mechanism gives rise to a rich spectrum of
potentially observable overtones, having periods of the order of
minutes, it also limits the modes to tiny amplitudes, typically
several parts-per-million in brightness.

The opportunity to continue asteroseismic studies of solar-type stars
with K2 would provide fresh data on stars in the solar neighborhood
for application to both stellar and Galactic chemical evolution
studies. The new fields have also led to the possibility of detecting
oscillations of solar-type stars in open clusters and eclipsing
binaries. This would provide independent data to test the accuracy of
asteroseismic estimates of fundamental stellar properties. Other
specific targets of interest would potentially benefit from the
provision of asteroseismic data, known exoplanet host stars being
obvious examples.

In this paper we report the detection of oscillations in several
subgiants using K2 short-cadence (SC) data collected during
Campaign\,1 (C1). We describe the target selection and data analysis,
and also discuss the implications of our results for future K2
campaigns.

 \section{Data}
 \label{sec:data}

 \subsection{Target selection and follow-up spectroscopic data}
 \label{sec:targetspec} 

Our selected target list started with the Hipparcos Catalog (van
Leeuwen 2007).  Use of these data allows us to make robust predictions
for many bright, potential K2 targets in the ecliptic.  Effective
temperatures were estimated from the $B-V$ color data in the catalog,
using the calibration of Casagrande et al. (2010), and luminosities,
$L$, were estimated from the parallaxes. These calculations used
reddening estimates from Drimmel et al. (2003) (negligible for many of
our targets).  We adopted $M_{\rm bol,\odot} = 4.73\,\rm mag$ (Torres
2010), and consistent bolometric corrections from the Flower (1996)
polynomials presented in Torres (2010), which use the estimated
$T_{\rm eff}$ as input. We also applied a cut on parallax, selecting
only those stars having fractional parallax uncertainties of 15\,\% or
better. Stellar radii were then estimated from $L$ and $T_{\rm eff}$,
and approximate masses were estimated from a simple power law in $L$
(which was sufficient for selecting targets).

%%%%%%%%%%%%%%%%%%%%%%%%%%%%%%%%%%%%%%%%%%%%%%%%%%%%%%%%%%%%%%%%%%%%%%%

\begin{figure*}
\epsscale{0.70}

\plotone{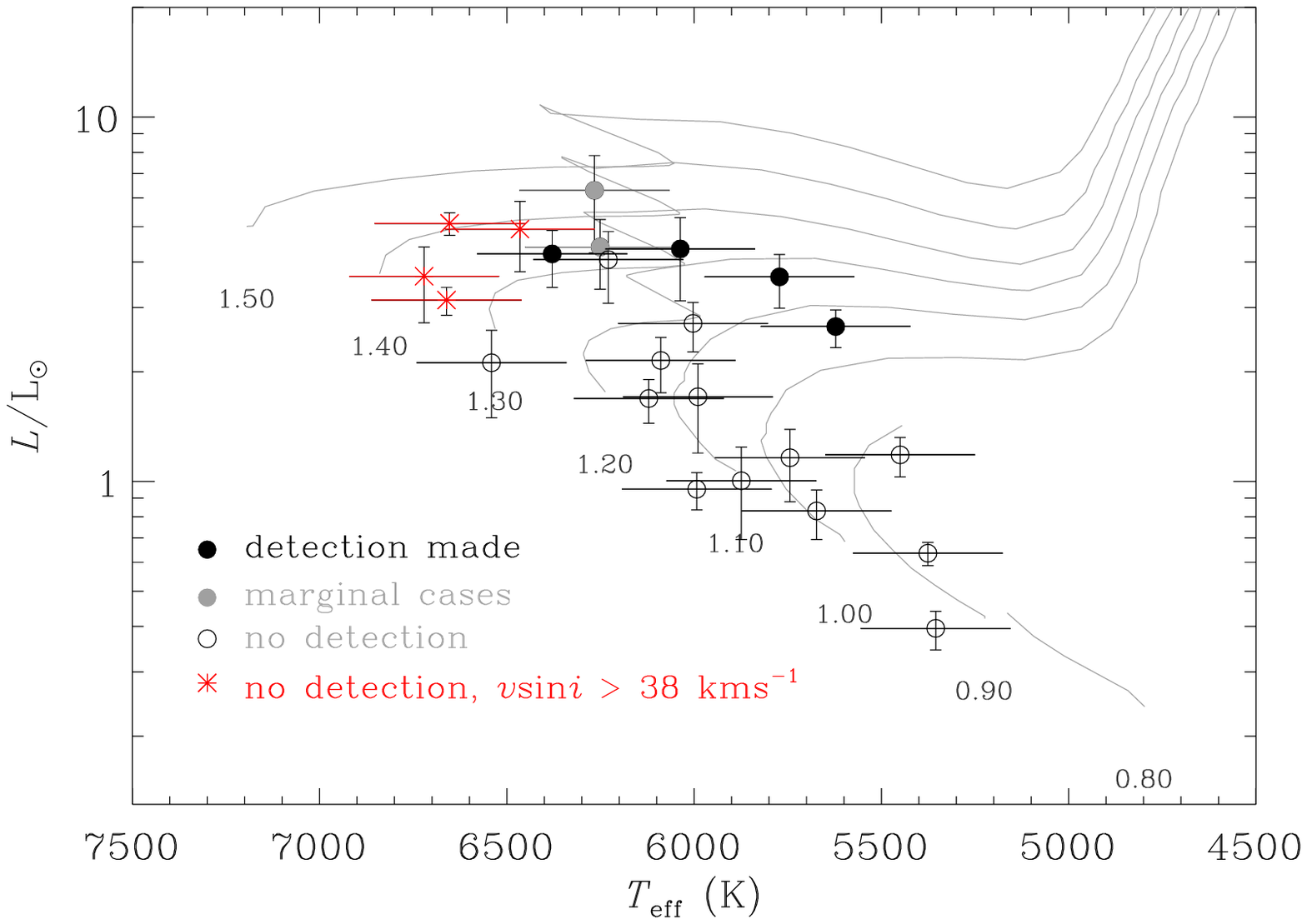}

\plotone{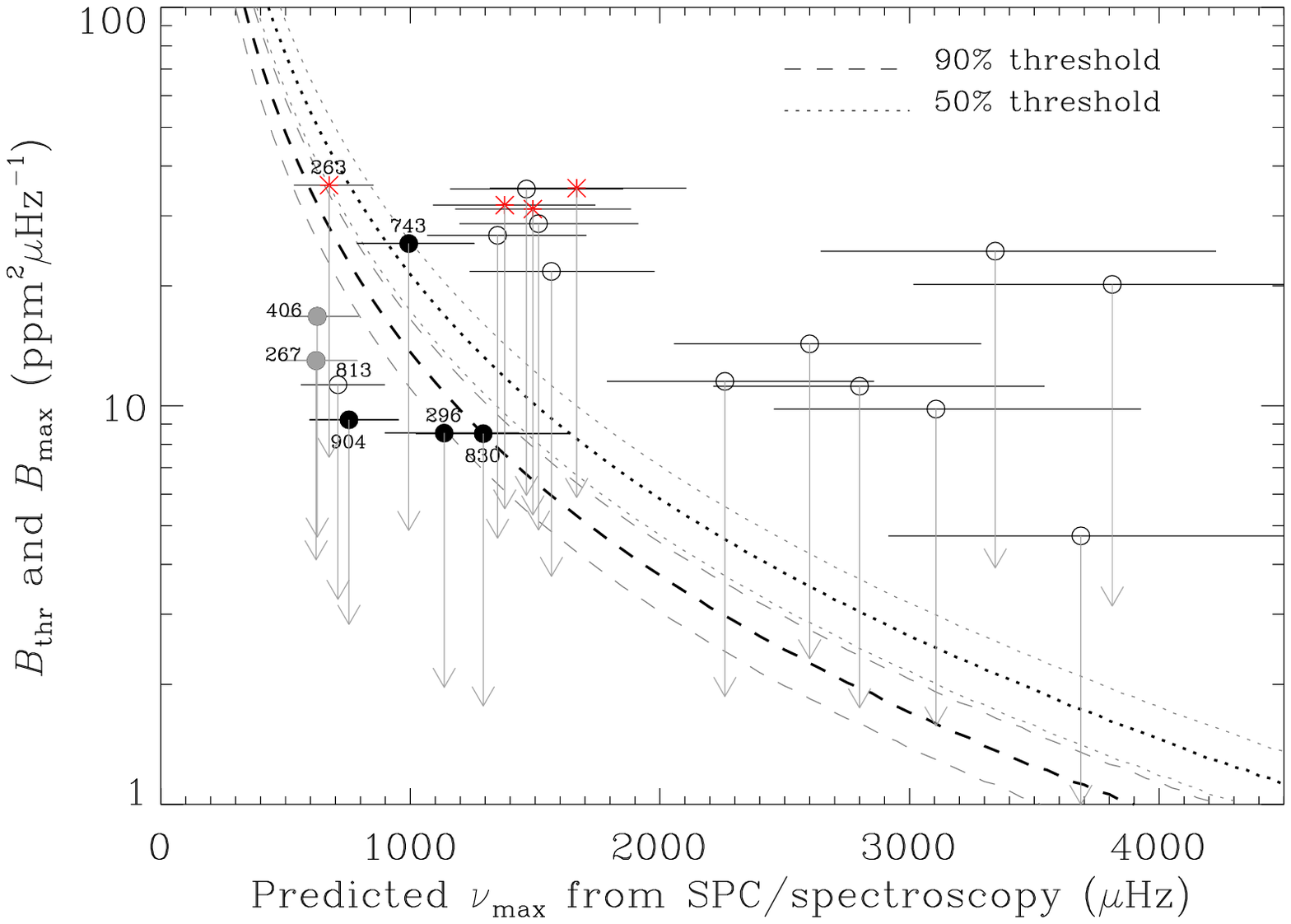}

\caption{Top panel: HR diagram of the SC targets.  Targets with firm
  asteroseismic detections are plotted with filled black circles;
  marginal detections with filled gray circles; no detections with
  open circles; and no detections with very high measured $v \sin i$
  (${\simeq 38}$ to $85\,\rm km\,s^{-1}$) with red asterisks. Bottom
  panel: Symbols with errors show measured background levels $B_{\rm
    max}$ in the K2 spectra (in power spectral density units) at the
  predicted $\nu_{\rm max}$, with rendering as per the top
  panel. Curves show threshold background power levels, $B_{\rm thr}$,
  to have a ${\simeq 90}\,\%$ (black dashed) and ${\simeq 50}\,\%$
  (black dotted) chance of making a detection. Uncertainties on the
  plotted thresholds are rendered in gray. Arrows show the result of
  reducing the excess high-frequency noise by a factor of
  two-and-a-half in amplitude (just over six in power).}

\label{fig:fig1}
\end{figure*}

%%%%%%%%%%%%%%%%%%%%%%%%%%%%%%%%%%%%%%%%%%%%%%%%%%%%%%%%%%%%%%%%%%%%%%%

The estimated Hipparcos-based stellar properties were used as input to
well-tested procedures (Chaplin et al. 2011b) that enabled us to
predict seismic parameters and relevant detectability metrics. We
narrowed down the sample to 23 well-characterized bright ($m_v \approx
7$ to 9) solar-type stars to be proposed for K2 observations. All
targets were predicted to show solar-like oscillations on timescales
of the order of minutes, necessitating SC observations.

We also collected ground-based spectroscopic data on our selected C1
targets to help us check the Hipparcos-based predictions, and to
better understand the final yield of asteroseismic detections.
Observations were made using the TRES spectrograph (F\"ur\'esz 2008)
on the 1.5-m Tillinghast telescope at the F. L. Whipple
Observatory. Atmospheric parameters were derived using the Stellar
Parameter Classification pipeline (SPC; see Buchhave et al. 2012). SPC
was used to match observed spectra -- taken at a resolution of 44000
-- to sets of synthetic model spectra to derive estimates of $T_{\rm
  eff}$, $\log\,g$, metallicity, and $v\sin i$. In what follows we
assume that relative metal abundances [m/H] returned by SPC are
equivalent to relative iron abundances, [Fe/H]. Table~\ref{tab:tab1}
contains the derived spectroscopic parameters.  There are four rapidly
rotating stars in the sample, and some caution is advised regarding
their estimated parameters. Overall, we found good agreement between
the spectroscopic parameters and the Hipparcos-based
values. Table~\ref{tab:tab1} also includes the Hipparcos-based
estimates of the luminosities.

To understand the limits on K2 performance in C1, we deliberately
sampled the region of the HR diagram across which detections had been
made in the nominal mission, as shown in the top panel of
Fig.~\ref{fig:fig1}. The symbols denote stars that provided firm
asteroseismic detections (black), marginal detections (gray), no
detections (open) or no detections with a high measured $v\sin i$ (red
asterisks). Details are given below (notably in
Section~\ref{sec:det}).

%%%%%%%%%%%%%%%%%%%%%%%%%%%%%%%%%%%%%%%%%%%%%%%%%%%%%%%%%%%%%%%%%%%%%%%%%%%%%%%%%%%%%%%

\begin{deluxetable}{llcccccc}
\tabletypesize{\tiny} \tablecaption{Spectroscopic parameters and luminosities of targets}
\tablewidth{0pt}
\tablehead{
 \colhead{EPIC}&
 \colhead{HIP}& 
 \colhead{$T_{\rm eff}$ ($\pm 77$)}&
 \colhead{$\log\,g$ ($\pm 0.1$)}&
 \colhead{[Fe/H] ($\pm 0.1$)}&
 \colhead{$v\sin i$ ($\pm 0.5$)}&
 \colhead{Predicted}&
 \colhead{$\log(L/\rm L_{\odot})$}\\
 \colhead{}&
 \colhead{}&
 \colhead{(K)}&
 \colhead{(dex)}&
 \colhead{(dex)}&
 \colhead{($\rm km\,s^{-1}$)}&
 \colhead{$\nu_{\rm max}$ ($\rm \mu Hz$)}&
 \colhead{(dex)}}
 \startdata
201162999&   56884& $5749$&  $4.44$&  $-0.13$& $  2.1$& $3106^{+ 821}_{- 649}$& $-0.08 \pm 0.07$\\
 & & & & & & & \\
201164031&   56907& $5723$&  $4.47$&  $ 0.38$& $  0.9$& $3343^{+ 884}_{- 699}$& $ 0.07 \pm 0.05$\\
 & & & & & & & \\
201182789\tablenotemark{1}&   57275& $6532$&  $4.15$&  $-0.04$& $ 38.4$& $1491^{+ 393}_{- 311}$& $ 0.69 \pm 0.09$\\
 & & & & & & & \\
201215315\tablenotemark{1}&   57456& $6523$&  $4.20$&  $-0.14$& $ 41.3$& $1666^{+ 439}_{- 347}$& $ 0.50 \pm 0.04$\\
 & & & & & & & \\
201343968&   55379& $6219$&  $4.09$&  $ 0.04$& $  9.0$& $1349^{+ 356}_{- 281}$& $ 0.61 \pm 0.09$\\
 & & & & & & & \\
201353392&   55288& $6110$&  $4.13$&  $-0.01$& $  8.1$& $1465^{+ 387}_{- 306}$& $ 0.43 \pm 0.07$\\
 & & & & & & & \\
201367296&   58093& $5695$&  $4.00$&  $ 0.19$& $  3.5$& $1135^{+ 300}_{- 237}$& $ 0.42 \pm 0.05$\\
 & & & & & & & \\
201367904&   58191& $6125$&  $3.84$&  $-0.04$& $ 10.3$& $ 754^{+ 199}_{- 157}$& $ 0.64 \pm 0.11$\\
 & & & & & & & \\
201421619&   55438& $5751$&  $4.39$&  $-0.38$& $  2.5$& $2799^{+ 740}_{- 585}$& $ 0.00 \pm 0.12$\\
 & & & & & & & \\
201436411&   56282& $6009$&  $4.15$&  $-0.26$& $  4.7$& $1565^{+ 413}_{- 327}$& $ 0.23 \pm 0.06$\\
 & & & & & & & \\
201592408&   56755& $5993$&  $4.31$&  $-0.24$& $  3.6$& $2260^{+ 597}_{- 472}$& $ 0.23 \pm 0.11$\\
 & & & & & & & \\
201601162&   54675& $5911$&  $4.53$&  $ 0.04$& $  5.6$& $3812^{+1007}_{- 796}$& $ 0.07 \pm 0.09$\\
 & & & & & & & \\
201602813&   55022& $6156$&  $3.81$&  $-0.82$& $ 11.9$& $ 710^{+ 187}_{- 148}$& $ 0.33 \pm 0.11$\\
 & & & & & & & \\
201614568\tablenotemark{1}&   54857& $6940$&  $4.13$&  $-0.05$& $ 85.8$& $1378^{+ 363}_{- 287}$& $ 0.71 \pm 0.03$\\
 & & & & & & & \\
201620616&   58643& $5999$&  $4.37$&  $-0.16$& $  3.6$& $2599^{+ 686}_{- 543}$& $-0.02 \pm 0.05$\\
 & & & & & & & \\
201626704&   54541& $5505$&  $4.50$&  $ 0.09$& $  0.7$& $3686^{+ 975}_{- 771}$& $-0.20 \pm 0.03$\\
 & & & & & & & \\
201698809&   55638& $5570$&  $4.71$&  $ 0.09$& $  0.4$& $5916^{+1565}_{-1237}$& $-0.40 \pm 0.05$\\
 & & & & & & & \\
201729267&   55574& $6130$&  $3.76$&  $-0.43$& $  7.0$& $ 622^{+ 164}_{- 130}$& $ 0.80 \pm 0.12$\\
 & & & & & & & \\
201733406&   55467& $6155$&  $3.76$&  $-0.29$& $  5.3$& $ 627^{+ 165}_{- 131}$& $ 0.64 \pm 0.09$\\
 & & & & & & & \\
201756263\tablenotemark{1}&   57034& $6820$&  $3.81$&  $-0.08$& $ 54.0$& $ 674^{+ 177}_{- 140}$& $ 0.56 \pm 0.10$\\
 & & & & & & & \\
201820830&   55778& $6417$&  $4.08$&  $ 0.00$& $ 12.2$& $1292^{+ 340}_{- 269}$& $ 0.62 \pm 0.08$\\
 & & & & & & & \\
201853942&   57136& $6053$&  $4.14$&  $-0.05$& $  4.4$& $1513^{+ 399}_{- 316}$& $ 0.33 \pm 0.07$\\
 & & & & & & & \\
201860743&   57676& $5852$&  $3.95$&  $-0.08$& $  4.7$& $ 993^{+ 262}_{- 207}$& $ 0.56 \pm 0.07$\\
 & & & & & & & \\
 \enddata
 \label{tab:tab1}

\tablenotetext{1}{Extra caution is advised regarding the
  classifications of these rapidly rotating stars.}
\end{deluxetable}

%%%%%%%%%%%%%%%%%%%%%%%%%%%%%%%%%%%%%%%%%%%%%%%%%%%%%%%%%%%%%%%%%%%%%%%%%%%%%%%%%%%%%%%

 \subsection{K2 Data and Lightcurve Preparation}
 \label{sec:light}

Each target was observed by K2 for just over 82\,days in C1---which
lasted from 2014 May 30 to 2014 August 21---with data collected in SC
mode (Gilliland et al. 2010b). We used the K2P$^2$ pipeline (Lund et
al. 2015) to prepare SC lightcurves for asteroseismic analysis. In
brief, the pipeline took the SC target pixel data as input. Masks for
all targets in a given frame were defined manually, and flux and
position data were then extracted for our chosen targets of
interest. Corrections were then applied to the lightcurves to mitigate
the impact of changes of the target positions on the CCD.  Finally,
additional corrections were made using the filtering described by
Handberg \& Lund (2014).

 \section{Results}
 \label{sec:res}

 \subsection{Asteroseismic Detections}
 \label{sec:det} 

The lightcurves were distributed to several teams, who each attempted
to detect signatures of solar-like oscillations in the power spectra
of the data. A complementary range of well-tested analysis codes was
used, which had been applied extensively to data from the nominal
\emph{Kepler} mission (e.g. Christensen-Dalsgaard et al. 2008; Huber
et al. 2009; Mosser \& Appourchaux 2009; Roxburgh 2009; Hekker et
al. 2010; Kallinger et al. 2010; Mathur et al. 2010; Gilliland et
al. 2011; Benomar et al. 2012; Campante 2012; see also comparison of
methods in Verner et al. 2011).  In cases where oscillations were
detected, each team was asked to return estimates of the two most
commonly used global or average asteroseismic parameters: $\Delta\nu$,
the average frequency spacing between consecutive overtones of the
same angular degree; and $\nu_{\rm max}$, the frequency at which the
oscillations present their strongest observed amplitudes.

We checked the asteroseismic detection yield using the spectroscopic
data.  The bottom panel of Fig.~\ref{fig:fig1} provides a visual
summary of these checks. The horizontal axis shows the predicted
$\nu_{\rm max}$ for each target, made using the spectroscopic $T_{\rm
  eff}$ and $\log g$ as input. Estimates were calculated using the
widely-used scaling relation (Brown et al. 1991, Kjeldsen \& Bedding
1995):
  \begin{equation} 
  \nu_{\rm max} \simeq \left(\frac{g}{\rm g_{\odot}}\right)
  \left(\frac{T_{\rm eff}}{\rm T_{\rm eff\,\odot}}\right)^{-1/2}\,
  \nu_{\rm max\,\odot},
  \label{eq:numax}
  \end{equation}
with the solar value $\nu_{\rm max\,\odot} = 3090\,\rm \mu Hz$ (see
Chaplin et al. 2014) providing the absolute calibration. These
predicted $\nu_{\rm max}$ are given in Table~\ref{tab:tab1}.

The vertical axis on the bottom panel of Fig.~\ref{fig:fig1} relates
to our ability to detect solar-like oscillations.  In the frequency
domain, where the analysis of the K2 data is conducted, peaks due to
the oscillations are superimposed on a slowly varying, broad-band
background. Our ability to make a detection depends on the prominence
of the oscillation peaks above that underlying background. For
solar-like oscillators, the background in the frequency range occupied
by the most prominent modes has contributions from granulation, shot
noise, and other instrumental noise.

The symbols with error bars show measured background levels, $B_{\rm
  max}$, in the K2 spectra at the predicted $\nu_{\rm
  max}$. Solar-like oscillators with lower frequencies of maximum
variability show larger amplitude oscillations: the lower $\nu_{\rm
  max}$, the larger is the maximum amplitude (i.e., brightness
variation, in ppm) and hence the easier it is to make a detection for
a given $B_{\rm max}$. Using the results from over 600 stars from the
nominal mission we can predict oscillation amplitudes as a function of
$\nu_{\rm max}$. With the predicted amplitudes in hand, we may
estimate threshold background levels that would permit a significant
detection of the oscillations.

We calculated these thresholds using the detection recipe in Chaplin
et al. (2011b). The lines in the bottom panel of Fig.~\ref{fig:fig1}
show the threshold levels, $B_{\rm thr}$, below which the observed
backgrounds $B_{\rm max}$ must lie to have a ${\simeq 90}\,\%$ (black
dashed) and ${\simeq 50}\,\%$ (black dotted) chance of making a
detection. The gray curves show uncertainties on the plotted
thresholds.

The results returned by the mode-detection teams indicated that we had
four good asteroseismic detections. These stars are plotted in
Fig.~\ref{fig:fig1} using filled black symbols; numbers are the final
three digits of the associated K2 Ecliptic Plane Catalog
(EPIC\footnote{http://archive.stsci.edu/k2/epic.pdf}) numbers
  (Huber et al., in preparation). Two other targets showed marginal
  detections, and they are shown in gray.  It is important to stress
  that these stars all lie in the part of the bottom panel where we
  would expect to make detections, i.e., where $B_{\rm max} < B_{\rm
    thr}$. Even though there are other targets that show lower or
  similar background levels in their K2 spectra, they lie at higher
  predicted $\nu_{\rm max}$ where the intrinsic oscillation
  amplitudes, and hence the chances of making a detection, are lower.

The observed $\nu_{\rm max}$ were in good agreement with the
spectroscopic predictions, at the level of precision of the data. Note
that the spectroscopic predictions of $\nu_{\rm max}$ have fractional
uncertainties of ${\simeq 25}\,\%$, which are significantly larger
than the typical uncertainties given by the asteroseismic measurements
(which are in contrast at the few-percent level).

%%%%%%%%%%%%%%%%%%%%%%%%%%%%%%%%%%%%%%%%%%%%%%%%%%%%%%%%%%%%%%%%%%%%%%%

\begin{figure*}
\epsscale{1.1}

\plottwo{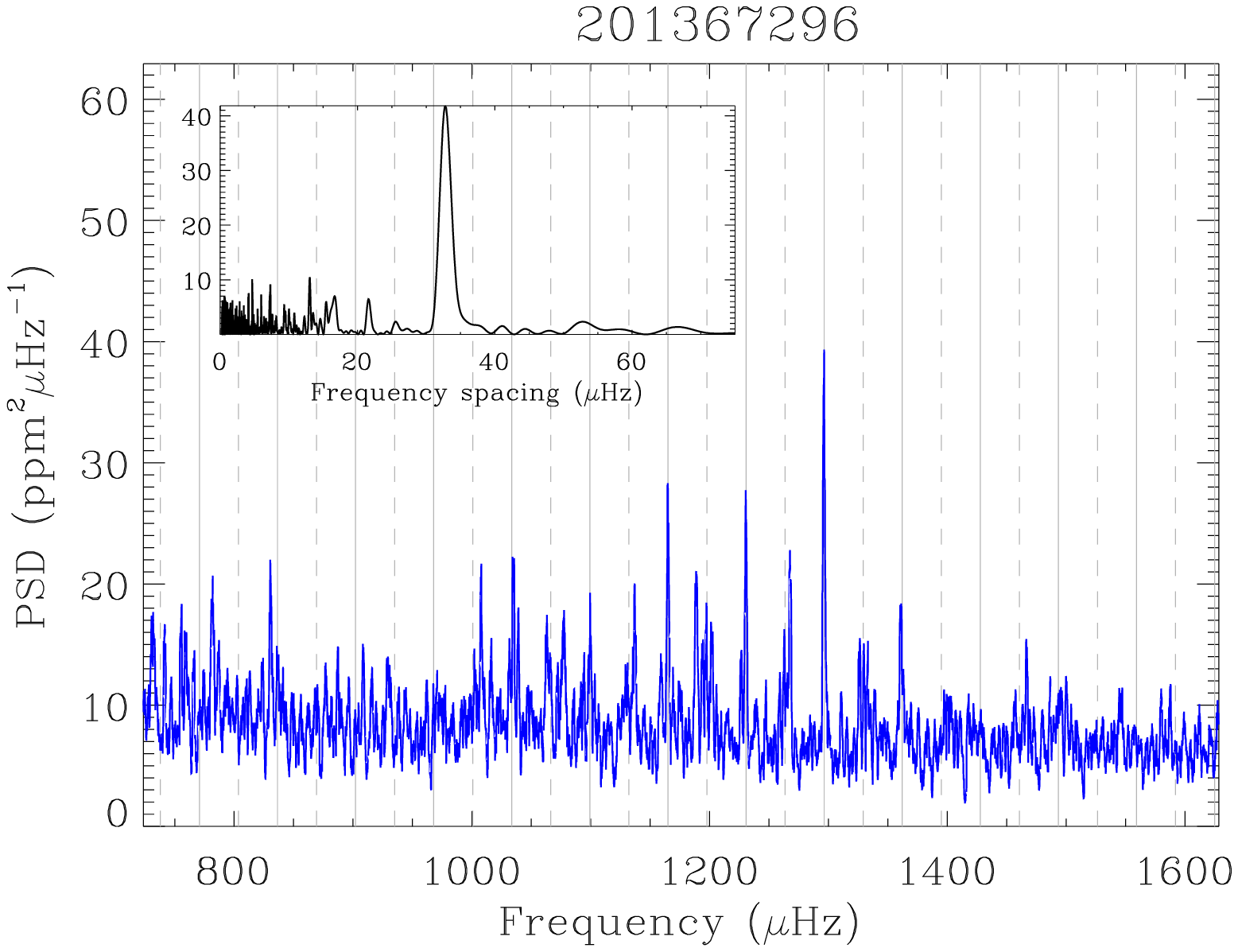}{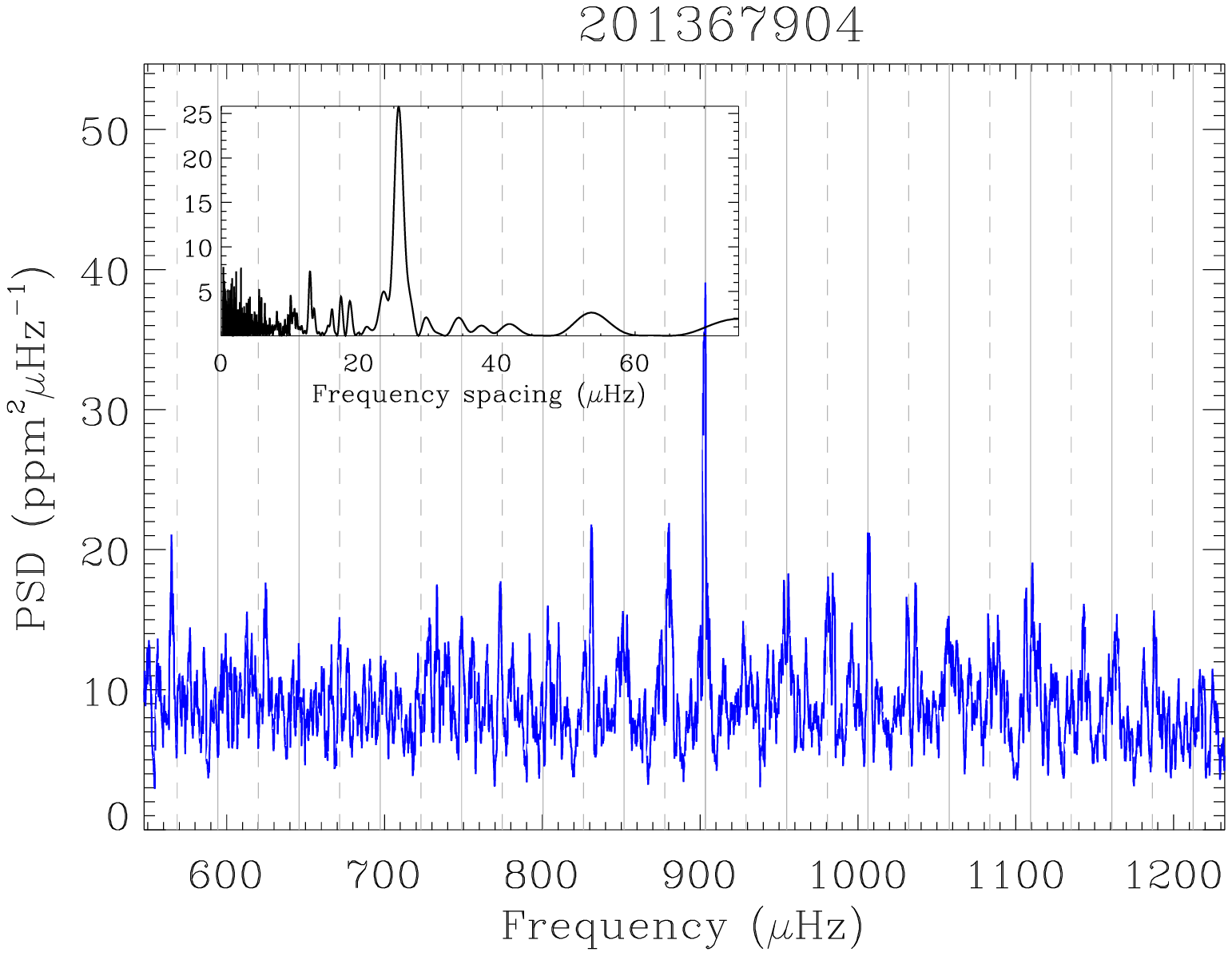}

\plottwo{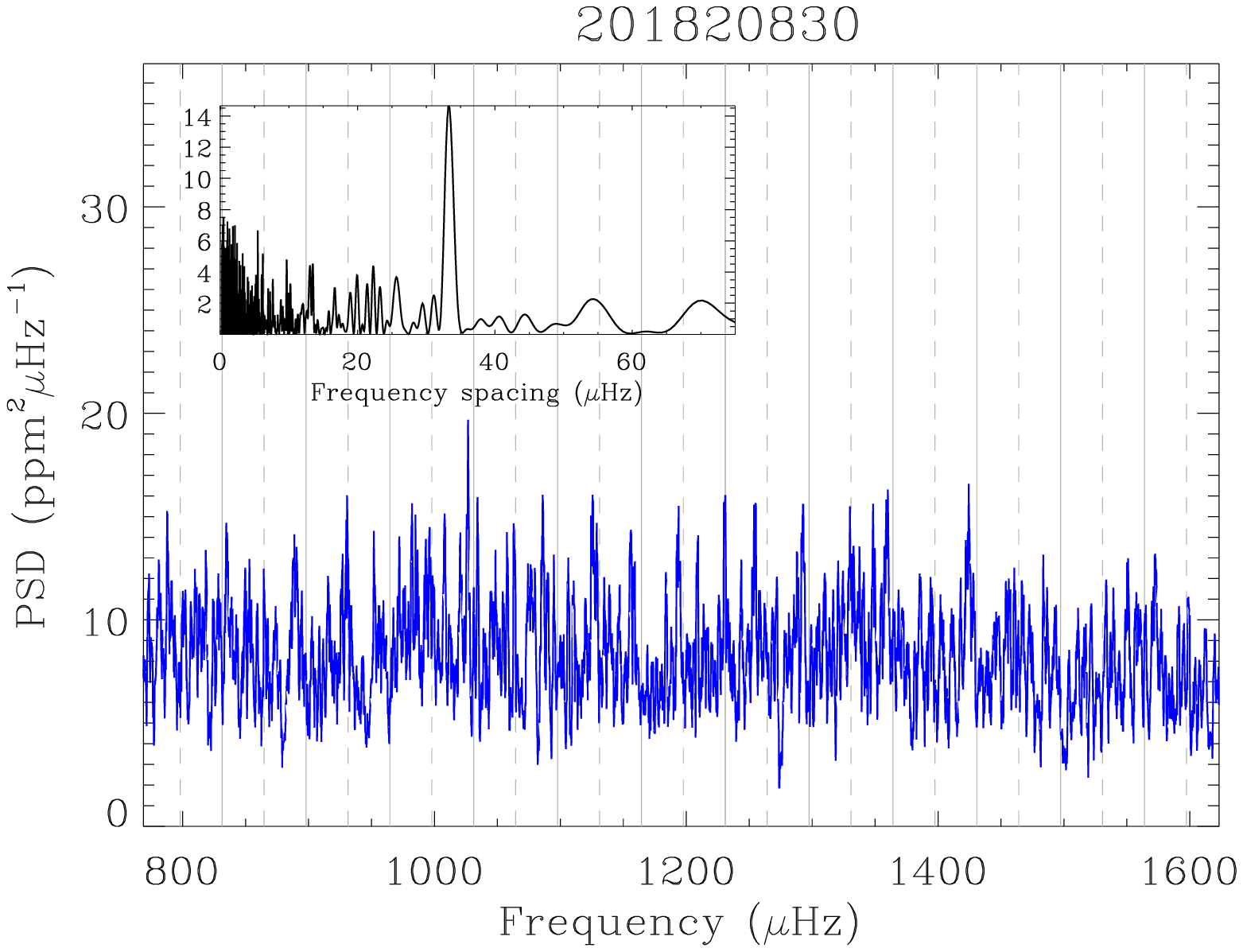}{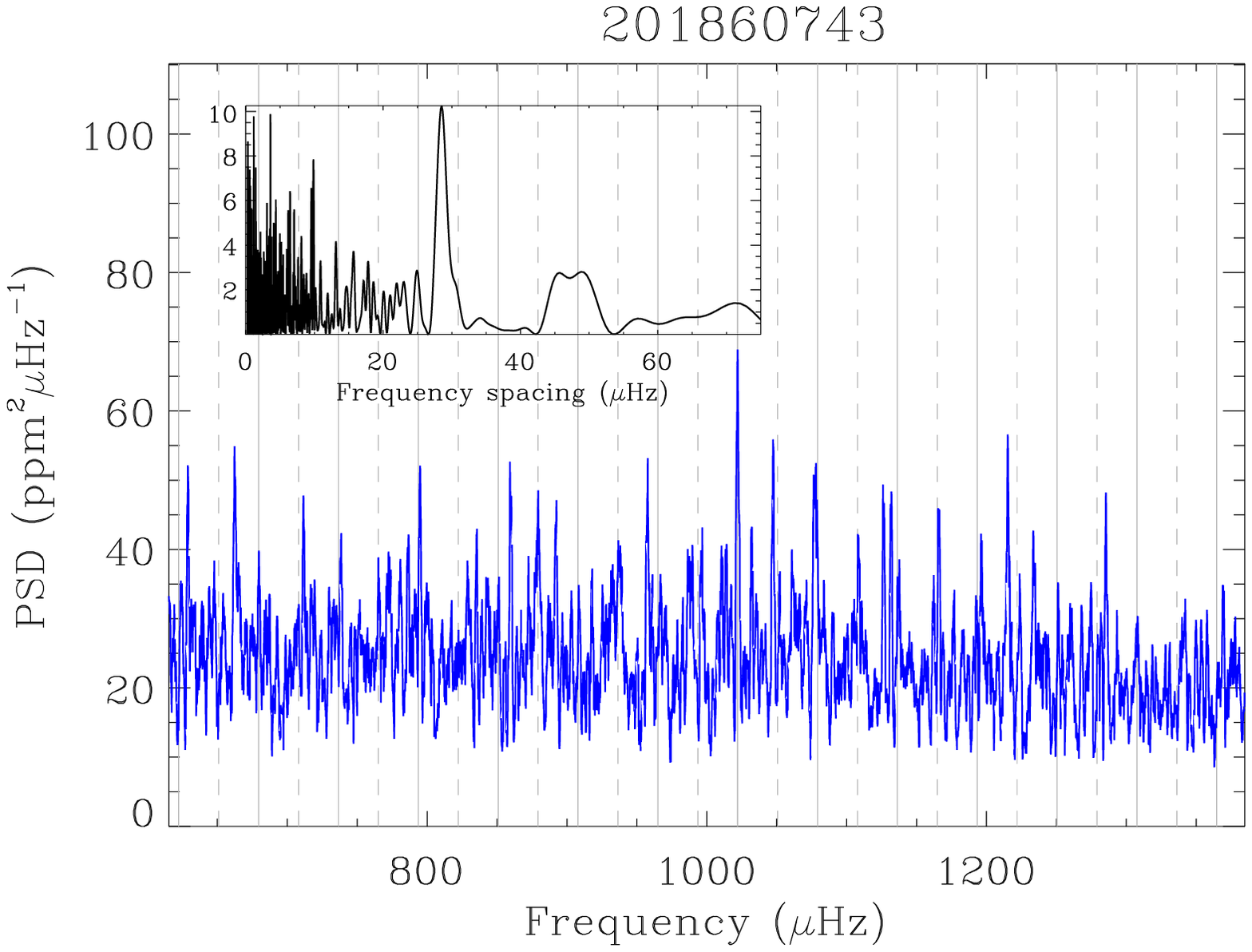}

\caption{K2 power spectra of the four targets with firm asteroseismic
  detections (smoothed with a $2\,\rm \mu Hz$ boxcar filter). Insets
  show the PSPS of each star, computed for regions in the first
  spectra that lie about $\nu_{\rm max}$. See text for further
  details.}

\label{fig:fig2}
\end{figure*}

%%%%%%%%%%%%%%%%%%%%%%%%%%%%%%%%%%%%%%%%%%%%%%%%%%%%%%%%%%%%%%%%%%%%%%%

Fig.~\ref{fig:fig2} shows the K2 power spectra of the four stars with
firm detections. Each spectrum has been smoothed with a $2\,\rm \mu
Hz$ boxcar filter. Sets of vertical gray solid and dashed lines are
separated by the estimated average $\Delta\nu$, and mark the spacing
on which we would expect to see modes.  The power envelope of the
oscillation spectrum of EPIC\,201820830 is somewhat flatter in
frequency than the more classic Gaussian-like envelopes shown by the
other three stars. This is a characteristic of hot F-type stars (e.g.,
Arentoft et al. 2008). The oscillations are strongly damped and this
tends to wash out the visual appearance of the oscillation spectrum.

The insets show the power spectrum of the power spectrum (PSPS) of
each star, computed from the region around $\nu_{\rm max}$. The
significant peak in each PSPS lies at $\Delta\nu/2$, and is the
detected signature of the near-regular spacing of oscillation peaks in
the frequency spectrum. These detection signatures persisted for each
star when the respective C1 lightcurves were divided into two equal
halves in the time domain and analyzed separately.

There are two stars in the the bottom panel of Fig.~\ref{fig:fig1}
that did not yield firm or even marginal detections, but might be
expected to do so. One of these stars (EPIC\,201756263) is plotted in
red. There are also three other stars shown in red on the
diagram. These are all rapidly rotating solar-type stars with $v\sin
i$ in the range ${\simeq 38}$ to $85\,\rm km\,s^{-1}$. They are
therefore presumably young and may be very active, which is known to
lead to significant suppression of the oscillation amplitudes (e.g.,
Garc\'ia et al. 2010; Chaplin et al. 2011c; Campante et al. 2014)
making it much harder to detect oscillations.  The detection recipe
does not yet make allowance for suppression of the oscillation
amplitudes by activity.

The other star that might be expected to show a detection
(EPIC\,201602813, plotted with an open symbol) is noteworthy because
it is by far the most metal-poor star in the sample. This may lead to
attenuation of the observed amplitudes, relative to the basic
predictions (e.g., Houdek et al. 1999; Samadi et al. 2010). Other
explanations are that the $\nu_{\rm max}$ scaling underpredicts the
$\nu_{\rm max}$ value for this star; or that the spectroscopic
parameters are incorrect. We think that this latter explanation is
unlikely. The target-selection (Hipparcos-based) prediction for
$\nu_{\rm max}$ was in good agreement with the spectroscopic estimate.
The target is also the primary component of a single-lined
spectroscopic binary (Carney et al. 2001), which will be spatially
unresolved in the K2 pixel data.  However, flux contamination from the
secondary (a suspected white dwarf) is likely to be low.

It is of course also possible that residual artifacts and other
problems relating to the lightcurve extraction, preparation and
filtering may have prevented a good detection being made. There are
clearly persistent artifact peaks in some of the frequency
spectra. Moreover, the high-frequency noise levels for all stars are
not close to being shot-noise limited, unlike the nominal
\emph{Kepler} data. Analysis of the K2 SC spectra therefore demands
some degree of careful, manual scrutiny to be sure that a claimed
detection is not the result of a chance combination of noise peaks.

Not including the active target EPIC\,201756263, of the seven cases
for which we would hope to detect oscillations, there are firm
detections in four and marginal detections in two. This is a good
return given the challenges posed by the K2 photometry.

 \subsection{Grid-Modeling Results}
 \label{sec:grid}

We consolidated the $\Delta\nu$ and $\nu_{\rm max}$ estimates returned
by the mode detection teams to give final asteroseismic parameters
ready for asteroseismic modeling. The final parameters were those
returned by an updated version of the OCTAVE pipeline (Chaplin et al.,
in preparation; see also Hekker et al. 2010). Its estimates had the
smallest average deviation from the median estimates in a global
comparison made over all pipelines and all stars. Uncertainties on
each final parameter were given by adding (in quadrature): the formal
parameter uncertainty given by the chosen pipeline; the standard
deviation of the parameter estimates given by all other pipelines; and
a small contribution to account for uncertainties in the solar
reference values (which are employed in the grid-modeling; see Chaplin
et al. 2014 for further discussion). The final parameters are listed
in Table~\ref{tab:tab2}.

A second set of teams were then asked to independently apply grid-based
modeling to estimate fundamental properties of the four stars with
firm detections. These teams, like their mode-detection counterparts,
used codes that have been applied extensively to nominal-mission
\emph{Kepler} data (e.g., see: Stello et al. 2009; Basu et al. 2010;
Kallinger et al. 2010; Quirion et al. 2010; Gai et al. 2011; Bazot et
al. 2012; Creevey et al. 2013; Hekker et al. 2013; Lundkvist et
al. 2014; Miglio et al. 2013; Serenelli et al. 2013; Hekker \& Ball
2014; Rodrigues et al. 2014; Silva Aguirre et al. 2015). Further
details may also be found in Chaplin et al. (2014) and Pinsonneault et
al. (2014).

We tested the impact on the estimated stellar properties of using
different sets of inputs, i.e., a first set with \{$\Delta\nu$,
$\nu_{\rm max}$, $T_{\rm eff}$, [Fe/H]\}, a second set with
\{$\Delta\nu$, $T_{\rm eff}$, [Fe/H]\}, and further sets with the
parallax-based luminosities also included. We found very good
agreement between the properties given by the different sets, and by
the different pipelines. Results from the first two sets were
consistent with the independent luminosity estimates.  Using the
luminosities as an additional input constraint did not have a
significant impact on the results. The parallax uncertainties---which
range from ${\simeq 5\,\%}$ to 12\,\%---are too large to add anything
useful to the seismic constraints. The fact that the second set of
inputs provided results that were consistent with the estimated
$\nu_{\rm max}$ lends further confidence to the claimed detections and
suggests that, at least for these data, potential bias in the
$\nu_{\rm max}$ estimates arising from spurious noise peaks is not a
significant cause for concern.

Table~\ref{tab:tab2} gives final values for the estimated properties,
using the first set of inputs. The properties were calculated using
the BeSPP pipeline (Serenelli et al. 2013), which uses individual
model frequencies to calculate model predictions of $\Delta\nu$ for
comparison with the observations. The uncertainties include a
contribution from the scatter between pipelines (following the
procedure outlined above for the input seismic parameters; see also
the in-depth discussions in Chaplin et al. 2014).

These grid-modeling results demonstrate that K2 has returned solid
results that allow asteroseismic modeling to be performed on the
targets. Two cases here -- EPIC\,201367296 and EPIC\,201367904 -- will
be amenable to more in-depth modelling studies since it will be
possible to extract precise and robust individual frequencies of
several overtones of each star.

%%%%%%%%%%%%%%%%%%%%%%%%%%%%%%%%%%%%%%%%%%%%%%%%%%%%%%%%%%%%%%%%%%%%%%%%%%%%%%%%%%%%%%%

\begin{deluxetable}{cccccccc}
\tabletypesize{\tiny} \tablecaption{Asteroseismic parameters and estimated stellar properties}
\tablewidth{0pt}
\tablehead{
 \colhead{EPIC}&
 \colhead{HIP}&
 \colhead{$\nu_{\rm max}$}& 
 \colhead{$\Delta\nu$}& 
 \colhead{$M$}&
 \colhead{$R$}&
 \colhead{$\rho$}&
 \colhead{$\log\,g$}\\
 \colhead{}&
 \colhead{}&
 \colhead{($\rm \mu Hz$)}&
 \colhead{($\rm \mu Hz$)}&
 \colhead{($\rm M_{\odot}$)}&
 \colhead{($\rm R_{\odot}$)}&
 \colhead{($\rm g\,cm^{-3}$)}&
 \colhead{(dex)}}
 \startdata
201367296& 58093& $1176 \pm 58$& $65.7 \pm 0.7$&$ 1.14 \pm 0.05$& $ 1.71 \pm 0.03$& $0.323 \pm 0.006$& $4.032 \pm 0.008$\\
201367904& 58191& $ 890 \pm 46$& $51.5 \pm 1.0$&$ 1.28 \pm 0.05$& $ 2.08 \pm 0.04$& $0.202 \pm 0.006$& $3.912 \pm 0.009$\\
201820830& 55778& $1196 \pm 72$& $66.6 \pm 0.8$&$ 1.35 \pm 0.06$& $ 1.77 \pm 0.03$& $0.345 \pm 0.006$& $4.076 \pm 0.008$\\
201860743& 57676& $1000 \pm 46$& $57.1 \pm 1.3$&$ 1.14 \pm 0.05$& $ 1.87 \pm 0.04$& $0.246 \pm 0.007$& $3.952 \pm 0.010$
 \enddata
 \label{tab:tab2}
\end{deluxetable}

%%%%%%%%%%%%%%%%%%%%%%%%%%%%%%%%%%%%%%%%%%%%%%%%%%%%%%%%%%%%%%%%%%%%%%%%%%%%%%%%%%%%%%%

\section{Summary}
\label{sec:sum}

We analysed K2 short-cadence (SC) data for 23 solar-type stars
observed in C1. Of the seven targets where we would hope to detect
oscillations, there are firm asteroseismic detections in four cases,
and marginal detections in a further two. This represents a good
return, in spite of the challenges posed by the K2 photometry. In sum,
we have a very good understanding of the asteroseismic yield.

The results put us in a good position to hone target selections for
future campaigns. Current performance levels mean we can detect
oscillations in sub-giants, but not in main-sequence stars. Changes to
the operation of the fine-guidance sensors are expected to give
significant improvements in the high-frequency performance of K2 from
C3 onwards. The high-frequency noise is currently a crucial limitation
to making asteroseismic detections, in particular in main-sequence
stars. With reference to the bottom panel of Fig.~\ref{fig:fig1}, we
note that a reduction of excess high frequency noise by a factor of
two-and-a-half in amplitude (just over six in power) would lead to
consistent detections of oscillations in main sequence stars with
$\nu_{\rm max}$ as high as ${\simeq 2500}\,\rm \mu Hz$, as well as
converting marginal detection cases to ones for which detailed
modeling could be performed.

The prospects are therefore very encouraging. Solar-type stars in the
Pleiades and Hyades open clusters have already been observed by K2 in
SC during C4. More stars will be observed in SC during C5 in the open
clusters M44 and M67. There is also now clear potential to build up a
statistical sample of solar-type field stars in the ecliptic that have
good asteroseismic data, and to target specific stars of interest for
asteroseismic study such as bright eclipsing binaries and known
exoplanet host stars.

\acknowledgements Funding for this Discovery mission is provided by
NASA's Science Mission Directorate. The authors wish to thank the
entire \emph{Kepler} team, without whom these results would not be
possible.  We also thank all funding councils and agencies that have
supported the activities of KASC Working Group\,1.


\begin{thebibliography}

\bibitem{} Arentoft, T., Kjeldsen, H., Bedding, T. R., et al., 2008, 
  ApJ, 687, 1180

\bibitem{} Basu, S., Chaplin, W. J., Elsworth, Y., 2010, ApJ, 710,
  1596

\bibitem{} Buchhave, L. A., Latham, D. W., Johansen, A., et al. 2012,
  Nature, 486, 375

\bibitem{} Benomar, O., Baudin, F., Chaplin, W. J., Elsworth, Y.,
  Appourchaux, T., 2012, MNRAS, 420, 2178

\bibitem{} Brown, T. M., Gilliland, R. L., Noyes, R. W., Ramsey,
  L. W., 1991, ApJ, 368, 599

\bibitem{} Campante, T. L. 2012, PhD thesis, Universidade do Porto
  (arXiv: 1405.3145)

\bibitem{} Campante, T. L., Chaplin, W. J., Lund, M. N., et al. 2014,
  ApJ, 783, 123

\bibitem{} Carney, B. W., Latham, D. W., Laird, J. B., Grant, C. E.,
  Morse, J. A., 2001, AJ, 122, 3419

\noindent Casagrande, L., Ram\'irez, I., Mel\'endez, J., Bessell, M.,
Asplund, M., et al., 2010, A\&A, 512, A54

\bibitem{} Chaplin, W. J., Kjeldsen, H., Christensen-Dalsgaard, J., et
  al., 2011a, Science, 332, 213

\bibitem{} Chaplin, W. J., Kjeldsen, H., Bedding, T. R., et al.,
  2011b, ApJ, 732, 54

\bibitem{} Chaplin, W. J., Bedding, T. R., Bonanno, A., et al., 2011c,
  ApJ, 732, 5

\bibitem{} Chaplin, W. J., Miglio, A., 2013, ARA\&A, 51, 353

\bibitem{} Chaplin, W. J., Basu, S., Huber, D., et al., 2014, ApJS,
  210, 1

\bibitem{} Christensen-Dalsgaard, J., Arentoft, T., Brown, T. M.,
  Gilliland, R. L., Kjeldsen, H., Borucki, W. J., Koch, D., 2008,
  JPhCS, 118, 012039

\bibitem{} Creevey, O. L., The{\'e}venin, F., Basu, S. et al.,
2013, MNRAS, 431, 2419

\bibitem{} Drimmel, R., Cabrera-Lavers, A., L\'opez-Corredoira,
  M. 2003, A\&A, 409, 205

\bibitem{} Flower, P. J., 1996, ApJ, 469, 355

\bibitem{} F\"ur\'esz, G. 2008, PhD thesis, Univ. of Szeged

\bibitem{} Gai, N., Basu, S., Chaplin, W. J., Elsworth, Y., 2011,
  ApJ, 730, 63

\bibitem{} Garc\'ia, R, A., Mathur, S., Salabert, D., et al., 2010,
  Sci, 329, 1032

\bibitem{} Gilliland, R. L., Brown, T. M., Christensen-Dalsgaard, J.,
  et al. 2010a, PASP, 122, 131

\bibitem{} Gilliland, R. L., Jenkins, J. M., Borucki, W. J., et al.,
  2010b, ApJ, 713, 160L

\bibitem{} Gilliland, R. L., McCullough, P. R., Nelan, E. P., et
  al. 2011, ApJ, 726, 2

\bibitem{} Handberg, R., Lund, M. N., 2014, MNRAS, 445, 2698

\bibitem{} Hekker, S., Broomhall, A.-M., Chaplin, W. J., et al.,
2010, MNRAS, 402, 2049

\bibitem{} Hekker, S., Elsworth, Y., Mosser, B., et al. 2013, A\&A,
  556, 59

\bibitem{} Hekker, S., Ball, W. H., 2014, A\&A, 564, 105

\bibitem{} Houdek, G., Balmforth, N. J., Christensen-Dalsgaard, J.,
  Gough, D. O., 1999, A\&A, 351, 582

\bibitem{} Huber, D., Stello, D., Bedding, T. R., et al., 2009,
CoAst, 160, 74

\bibitem{} Huber, D., Chaplin, W. J., Christensen-Dalsgaard, J., et
  al., 2013, ApJ, 767, 127

\bibitem{} Howell, S. B., Sobeck, C., Haas, M., et al., 2014, PASP,
  126, 398

\bibitem{} Kallinger, T., Mosser, B., Hekker, S., et al. 2010, A\&A,
  522, 1

\bibitem{} Kjeldsen, H., Bedding, T. R., 1995, A\&A, 293, 87

\bibitem{} Lund, M. N., Handberg, R., Davies, G. R., Chaplin, W. J.,
  Jones, C. D., 2015, ApJ, 806, 30

\bibitem{} Lundkvist, M., Kjeldsen, H., Silva Aguirre, V., 2014, A\&A,
  566, 82

\bibitem{} Mathur, S., Garc\'\i a, R. A., R\'egulo C., et al., 2010,
  A\&A, 511, 46

\bibitem{} Miglio, A., Chiappini, C., Morel, T., et al., 2013,
  MNRAS, 429, 423

\bibitem{} Mosser, B., Appourchaux, T., 2009, A\&A, 508, 877

\bibitem{} Pinsonneault, M. H., Elsworth, Y., Epstein, C., et
  al. 2014, ApJS, 215, 19

\bibitem{} Quirion, P.-O., Christensen-Dalasgaard, J., Arentoft, T.,
  2010, ApJ, 725, 2176

\bibitem{} Rodrigues, T. S., Girardi, L., Miglio, A., et al.,
  2014, MNRAS, 445, 2758

\bibitem{} Roxburgh, I. W., 2009, A\&A, 506, 435

\bibitem{} Samadi, R., Ludwig, H.-G., Belkacem, K., Goupil, M. J.,
  Dupret, M.-A., 2010, A\&A, 509, 15

\bibitem{} Serenelli, A. M., Bergemann, M., Ruchti, G.,
  Casagrande, L., 2013, MNRAS, 429, 3645

\bibitem[]{} Silva Aguirre, V., Davies, G. R., Basu, S., et.
  al. 2015, MNRAS, in press (arXiv:1504.07992)

\bibitem[]{} Stello, D., Chaplin, W. J., Bruntt, H., et al. 2009,
  ApJ, 700, 1589

\bibitem{} Torres, G., 2010, AJ, 140, 1158

\bibitem[]{} Torres, G., Fischer, D. A., Sozzetti, A., Buchhave,
  L. A., Winn, J. N., Holman, M. J., Carter, J. A. 2012, ApJ, 757, 161

\bibitem{} van Leeuwen, F. 2007, A\&A, 474, 653

\bibitem{} Verner, G. A., Elsworth, Y., Chaplin, W. J., 2011, MNRAS,
  415, 3539


%%%%%%%%%%%%%%%%%%%%%%%%%%%%%%%%%%%%%%%%%%%%%%%%%%%%%%%%%%%%%%%%%%%%

\end{thebibliography}
\end{document}